\documentclass[trackchanges]{aastex701}
\usepackage{natbib}
\usepackage{graphicx, caption, subcaption}

\begin{document}

\title{Observation of Large-Scale Kelvin-Helmholtz Instability Wave Driven by a Coronal Mass Ejection}

\correspondingauthor{Leon Ofman}
\author[0000-0003-0602-6693]{Leon Ofman}
\affiliation{Institute for Astrophysics and Computational Sciences, Department of Physics, 
Catholic University of America, 
Washington, DC 20064, USA}
\affiliation{Heliophysics Science Division, 
NASA Goddard Space Flight Center, 
Greenbelt, MD 20771, USA}
\altaffiliation{Visiting, Department of Geosciences,  Tel Aviv University, Tel Aviv, Israel}
\email{show}{ofman@cua.edu}

\author[0000-0002-3230-2033]{Olga Khabarova}
\affiliation{Interdisciplinary Centre for Security, Reliability and Trust, University of Luxembourg, Luxembourg}
\email{olga.khabarova@uni.lu}

\author[0000-0002-2106-9168]{Ryun-Yong Kwon}
\affiliation{Korea Astronomy and Space Science Institute(KASI), Daejeon, S. Korea}
\email{rkwon@kasi.re.kr}

\author[0000-0001-6018-9018]{Yogesh}
\affiliation{Department of Physics and Astronomy, University of Iowa, Iowa City IA 54224, USA}
%\affiliation{NASA Goddard Space Flight Center, Greenbelt, MD, 20771, USA}
%\affiliation{The Catholic University of America, Washington, DC 20064, USA}
\email{yphy22@gmail.com}

\author[0000-0002-3584-3978]{Eyal Heifetz}
\affiliation{Department of Geosciences, Tel Aviv University, Ramat Aviv, Tel Aviv, Israel}
\email{eyalh@tauex.tau.ac.il}

\author[0000-0002-6905-9487]{Katariina Nykyri}
\affiliation{Embry-Riddle Aeronautical University, Daytona Beach, FL, USA}
\email{nykyrik@erau.edu}

%% Use the \collaboration command to identify collaborations. This command
%% takes an optional argument that is either a number or the word "all"
%% which tells the compiler how many of the authors above the command to
%% show. For example "\collaboration[all]{(DELVE Collaboration)}" wil include
%% all the authors above this command.
%%
%% Mark off the abstract in the ``abstract'' environment. 
\begin{abstract}

The Kelvin-Helmholtz instability (KHI) can occur when there is a relative motion between two adjacent fluids. In the case of magnetized plasma, the shear velocity must exceed the local Alfv\'{e}n speed for the instability to develop. The KHI produces nonlinear waves that eventually roll up into vortices and contribute to turbulence and dissipation.  In the solar atmosphere KHI  has been detected in coronal mass ejections (CMEs), jets, and prominences, mainly in the low corona. Only a few studies have reported the KHI in the upper corona, and its vortex development there has not been previously observed. We report the event with large-scale KHI waves observed from $\sim 6$ to 14~$R_{\odot}$ on 2024-Feb-16 using SOHO/LASCO and STEREO-A coronagraphs. KHI appeared during the passage of a fast CME and evolved into the nonlinear stage showing evidence of vortices. A closely timed subsequent CME in the same region has further developed the fully nonlinear KHI waves along its flank. We find that the radial speed of the CMEs exceeds the estimated local Alfv\'{e}n speed obtained from in-situ Parker Solar Probe (PSP) magnetic field data at perihelia. We propose that such events are rare because the fast CME created specific conditions favorable for instability growth in its trailing edge, including radial elongation of magnetic-field lines, reduced plasma density, and enhanced velocity and magnetic-field shear along the developing interface. The observed growth rate of KHI wave is in qualitative agreement with the theoretical predictions.
\end{abstract}

%% Keywords should appear after the \end{abstract} command. 
%% The AAS Journals now uses Unified Astronomy Thesaurus (UAT) concepts:
%% https://astrothesaurus.org
%% You will be asked to selected these concepts during the submission process
%% but this old "keyword" functionality is maintained in case authors want
%% to include these concepts in their preprints.
%%
%% You can use the \uat command to link your UAT concepts back its source.

%% From the front matter, we move on to the body of the paper.
%% Sections are demarcated by \section and \subsection, respectively.
%% Observe the use of the LaTeX \label
%% command after the \subsection to give a symbolic KEY to the
%% subsection for cross-referencing in a \ref command.
%% You can use LaTeX's \ref and \label commands to keep track of
%% cross-references to sections, equations, tables, and figures.
%% That way, if you change the order of any elements, LaTeX will
%% automatically renumber them.

\section{Introduction} 
\label{int:sec}
The  Coronal Mass Ejections (CMEs) are routinely observed with space-based coronagraphs such as LASCO on-board the Solar and Heliospheric Observatory \citep[SOHO;][]{Dom95}, and Sun Earth Connection Coronal and Heliospheric Investigation (SECCHI) COR 1 and COR 2 coronagraphs on the Solar TErrestrial RElations Observatory \citep[STEREO;][]{Kai08} spacecraft. During periods of increased solar activity the CMEs can reach speeds of several hundred to $\sim$2000 km s$^{-1}$, and form fast shocks with Alfv\'{e}nic Mach number $M_A>1$, i.e., with eruption velocities exceeding the local Alfv\'{e}n speed, $V_A$. These energetic eruptions, observed as they form in the solar corona, propagate throughout the heliosphere reaching the Earth's orbit and beyond. The CME's fast propagation results in the velocity shear between the nearly stationary background coronal structures, such as adjacent streamers, and this can lead to the development of shear-driven magnetized fluid instability as well as generation of streamer waves. Several studies have analyzed coronal streamer waves produced by CMEs \citep{Che10,Che11,Kwo13a,Kwo13b,Sri16,Dec20}, showing that the characteristics of these waves can be used to extract valuable seismological information, such as the magnetic field in both the streamers and the CME ejecta-sheath interface \citep{Nykyri2013}.

The Kelvin Helmholtz instability (KHI) can grow due to shear flows in magnetized fluids when the velocity shear (or velocity jump) magnitude exceeds the local Alfv\'{e}n speed in the parallel direction \citep{Cha61}. The magnetized KHI instability was previously detected in the Earth's magnetospheric boundary where it plays an important role in energy transport \citep[see the review,][]{Nyk21}. In the context of solar phenomena, KHI often arises due to differences in the plasma velocity configuration between adjacent regions or within magneto-plasma structures that move at different speeds relative to the surrounding plasmas. Several studies have documented the development of the KHI in solar prominences \citep{Ber10,Ber17,Hillier18,Yang18}, in coronal jets \citep{Li18,Zhe18,ZC18,Bog18,Mis25}, in CMEs \citep[e.g.,][]{OT11,Fou11,Mos13,Nyk24}, streamers \citep[e.g.,][]{Fen13}, and at the edges of coronal holes \citep{Telloni22}, switchback boundaries \citep{Moz20,Cho25}, and detected in-situ by the Solar Orbiter spacecraft in the solar wind \citep{Kie21}. Recently KHI signatures of small scale vortices were directly imaged close to the Sun in the range $7.5-9.5 R_s$ (where $R_s$ is the solar radius) with Parker Solar Probe (PSP) \citep{Fox16} Wide Field Imager (WISPR, \citet{Vou16}) in a slow CME ($\sim$ 200 km s$^{-1}$) that propagated along a preexisting streamer \citep{Pao24}. 

Understanding the development of KHI in the solar corona is important for explaining various solar phenomena, as it directly impacts energy transfer processes and the stability of the solar atmosphere. In particular, the KHI may lead to the formation of nonlinear waves that eventually break and create turbulence, which is, in turn, linked to corona heating, the most debated topic in the solar community \citep[see the review,][]{Van20}.

The present study is motivated by observations of a large amplitude wave in a dense coronal structure at distances $\sim$6 to 14 $R_s$ associated with  propagation of two fast CMEs subsequently ejected from the same active region detected on 2024 February 16 with COR1 and COR2 coronagraphs on board STEREO A and with the LASCO C2 and C3 coronagraphs on board SOHO.  This is the first observational analysis of the growth and saturation of large-amplitude nonlinear wave structure (vortex) that is a typical signature of KHI, and subsequent dissipation of this structure in this region.  The results impact the understanding of evolution of magnetized flows and CMEs in the lower corona, and the potentially important role  of KHI in the energy transfer in the solar corona in eruptive events. The paper is organized as follows. In Section~\ref{CME:sec} we analyze the CME evolution, and Section~\ref{okhi:sec} we present the analysis of the observed KHI wave, in Section~\ref{psp_av:sec} we present the PSP results of the coronal properties, Section~\ref{tkhi:sec} is devoted to the theoretical analysis, and the Discussion and Conclusions are in Section~\ref{disc:sec}. 

\section{Observational Data Analysis: CME Evolution} 
\label{CME:sec}

\begin{figure}[ht]
%\begin{center}
%  \begin{subfigure}{0.75\textwidth}
%    \includegraphics[width=1.0\textwidth]{STEREO-LASCO_v1.png}
 % \end{subfigure}
 %    \begin{subfigure}{0.75\textwidth}
 %    \hspace{-0.7cm} \includegraphics[width=0.6\textwidth]{CME_t-h_v2.pdf}
 %     \includegraphics[width=0.45\textwidth]{STEREO_position_v1.png}
  %\end{subfigure}
 \includegraphics[width=1.0\textwidth]{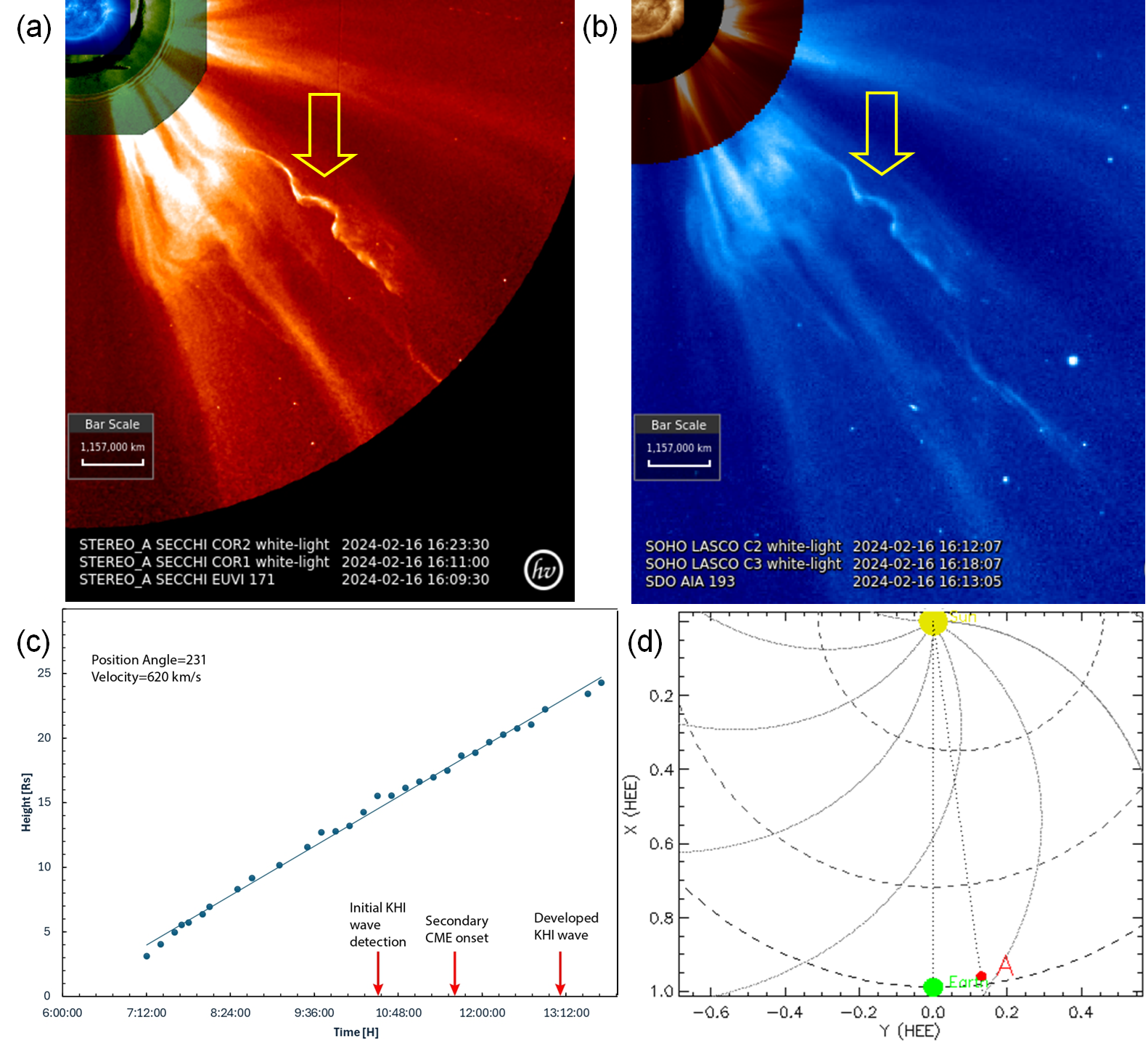}
\caption{The CME and the associated wave marked with the yellow arrow observed on 2024 February 16. The timing of the various instruments are indicated on the panels. (a) STEREO Cor 2 (the inner regions are obtained from EUVI and Cor 1 images). (b) The SOHO/LASCO C3 observations (the inner regions are obtained from LASCO C2, C3 and SDO AIA 193\AA\ images). The animations of this event produced by JHelioviewer (see, \url{https://www.jhelioviewer.org}) are available online. (c) Height-time plot of the main CME observed on 2024 February 16. The CME speed in the plane of the sky obtained from linear least square fit is 620 km s$^{-1}$, with $R^2=0.9959$. The red arrows indicate the approximate onset times of the second CME, initial wave formation,  and the observed developed KHI wave. (d) STEREO A view of the Sun (red dot) on 2024 February 16 with $\sim 8^\circ$ separation angle with respect to LASCO. }
\label{stereo-lasco:fig}
%\end{center}
\end{figure}

The fast partial-halo CME  on 2024 February 16 with onset time 7:12:05 was observed by STEREO A COR1 and COR2 coronagraphs, as well as by SOHO/LASCO C2 and C3 coronagraphs. The parameters of the CME and the animations are available on the SOHO LASCO CME catalog at \url{https://cdaw.gsfc.nasa.gov/CME_list}. In Figure~\ref{stereo-lasco:fig}a the CME and the associated wave marked with the yellow arrow are shown from STEREO COR1 data with the inner corona imaged with EUVI 171\AA\ emission. The developed KHI large amplitude wave structure extends nearly radially, marked by yellow arrow. Figure~\ref{stereo-lasco:fig}b show the LASCO C2 and C3 images with the inner corona from SDO AIA 193\AA\ emission.  The heliocentric distances in Figure~\ref{stereo-lasco:fig}a and b are matched to facilitate comparison. A large-amplitude KHI wave structure indicated by the yellow arrows in the upper panels extends nearly radially, likely associated with partially erupted prominence material. The nearly radial KHI large amplitude wave structure is also evident in LASCO C3 image. The heliocentric separation between the two spacecraft was $\sim8^\circ$, shown schematically in Figure~\ref{stereo-lasco:fig}c. The animations for this figure were produced using JHelioviewer software (see, \url{https://www.jhelioviewer.org}).

The well-developed KHI event on 16 February 2024 is associated with two consecutive CMEs originating from a broad active region located near S19W85, close to the southwestern solar limb. The primary CME ($v\approx620$ km$^{-1}$) was driven by an intense X2.5-class flare that occurred at 06:42 UT on 16 February 2024. According to the NASA GOES flare catalogue (\url{https://umbra.nascom.nasa.gov/goes/eventlists/goes_event_listings/goes_xray_event_list_2024.txt}), based on GOES X-ray flux measurements, the active region exhibited elevated flare activity beginning one day earlier, producing several C- and M-class flares; the X2.5-class flare associated with the event discussed here represented the peak of its eruptive evolution. A video of the flare sequence together with the corresponding GOES X-ray flux at Earth orbit is available in the SDO archive:
(\url{https://sdowww.lmsal.com/sdomedia/ssw/media/ssw/ssw_client/data/ssw_service_240216_202757_30626/www/EDS_FlareDetective-TriggerModule_20240214T131759-20240217T041235_AIA_211_XCEN883.2YCEN-268.8_ssw_cutout_20240214_131759_AIA_211__20240214_131757_m.mp4}

The manually compiled SOHO/LASCO CME catalogue maintained by the CDAW Data Center (\url{https://cdaw.gsfc.nasa.gov/CME_list/UNIVERSAL_ver2/2024_02/univ2024_02.html}) lists the first appearance of the main CME in coronagraph images at 2024/02/16 07:12:05 UT. A second, weaker and slower CME ($v\sim450$ km s$^{-1}$) erupted approximately five hours later (see,  Figure~\ref{stereo-lasco:fig}b animation and \url{https://www.sidc.be/cactus/}). Owing to its low brightness relative to the primary CME, this secondary eruption was not identified as a separate event in widely used CME catalogues and is therefore treated here as part of the main eruption sequence. The KHI initiated forming  waves and vortices during the propagation of the main CME, while  the subsequent CME  extends the duration of the fast outflow, contributing to further KHI wave growth.

%1

In Figure~\ref{stereo-lasco:fig}c the height-time plot for this CME is shown. The least square fit velocity is 620 km s$^{-1}$, while a second order polynomial produced slightly decelerating velocity profile resulting in about 670 km s$^{-1}$ at 10$R_s$, with both fits producing similar high correlation coefficient $R^2$ values near unity. The approximate onset time of the second CME and the detection time of developed KHI wave with vortices (see, Figure~\ref{slit_vr:fig} below) are indicated with the red arrows. We note that the CME front has traveled away from the lower heights of KHI wave growth with the CME front detected at $>20R_s$ at the indicated time of developed KHI wave. The KHI develops over the entire range of distances shown in Figure~\ref{stereo-lasco:fig}a-b, and the red arrows merely show the timing for the reader’s convenience.

\subsection{Observed KHI Wave Properties} 
\label{okhi:sec}

\begin{figure}[h]
%\begin{center}
%    \begin{subfigure}{0.66\textwidth}
      \includegraphics[width=\textwidth]{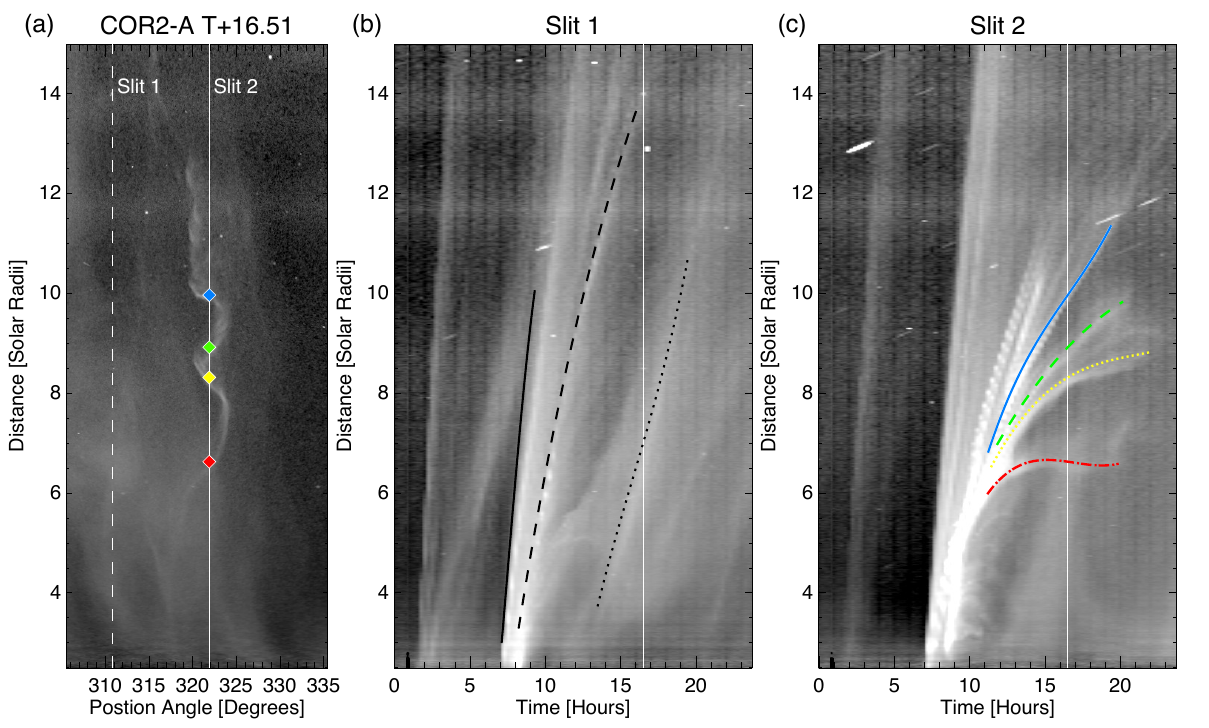}
%      \caption{The rectangle is a complicated geometrical figure that has 4 edges and 4 vertices while a star is an even more complex geometrical figure}
 %   \end{subfigure}
   % \begin{subfigure}[d]{0.34\textwidth}
 %     \includegraphics[width=\textwidth]{KH_Vortices.png}
%      \caption{A simple circle and a spiral}
  %  \end{subfigure}
%\includegraphics[width=0.66\textwidth]{slit_velocities_v2.pdf}
%\includegraphics[width=0.33\textwidth]{KH_Vortices.png}
\caption{(a) Portion of the STEREO COR2-A field of view at 16:38 UT on 2024 February 16, highlighting the KHI wave event. The two vertical dashed and solid lines refer to the two virtual slit positions used to construct the distance–time plots in panels (b) and (c). The position of Slit 1 (dashed line) is fixed, while Slit 2 (solid line) crosses the KHI wave features. Colored dots along Slit 2 mark the locations used to track the wave structure. (b) Distance–time plot of the propagating features along Slit 1, where features \#1 (solid), \#2 (dashed), and \#3 (dotted) are identified. (c) Distance–time plot of the KHI wave features along Slit 2, showing the evolution of features \#1 (blue), \#2 (green), \#3 (yellow), and \#4 (red), which correspond to the colored dots with the same color code. Colored curves indicate the fitted propagation tracks used to derive their velocities. The vertical solid line in panels (b) and (c) correspond to the time of the panel (a).  An animation of this figure is available online. }
\label{slit_vr:fig}
%\end{center}
\end{figure}

The observed KHI wave properties were obtained from the space-time analysis of the waves feature shown in Figure~\ref{stereo-lasco:fig}. In Figure~\ref{slit_vr:fig}a the wave feature is shown in a zoomed-in field of view, focusing on the wave structure with two virtual slits marked on the image with the vertical lines. The location of slit~1  is in an adjacent erupting region and is used to provide the space-time plot in Figure~\ref{slit_vr:fig}b. The location of slit~2 is chosen to cut through the wave features, that allows determining the time-depended wavelength (see the colored dots) and the velocities of the wave structure  from the space-time plots shown in Figure~\ref{slit_vr:fig}c. 
%2

%3
\begin{figure}[h]
     \hspace{0.5cm} \includegraphics[width=\textwidth]{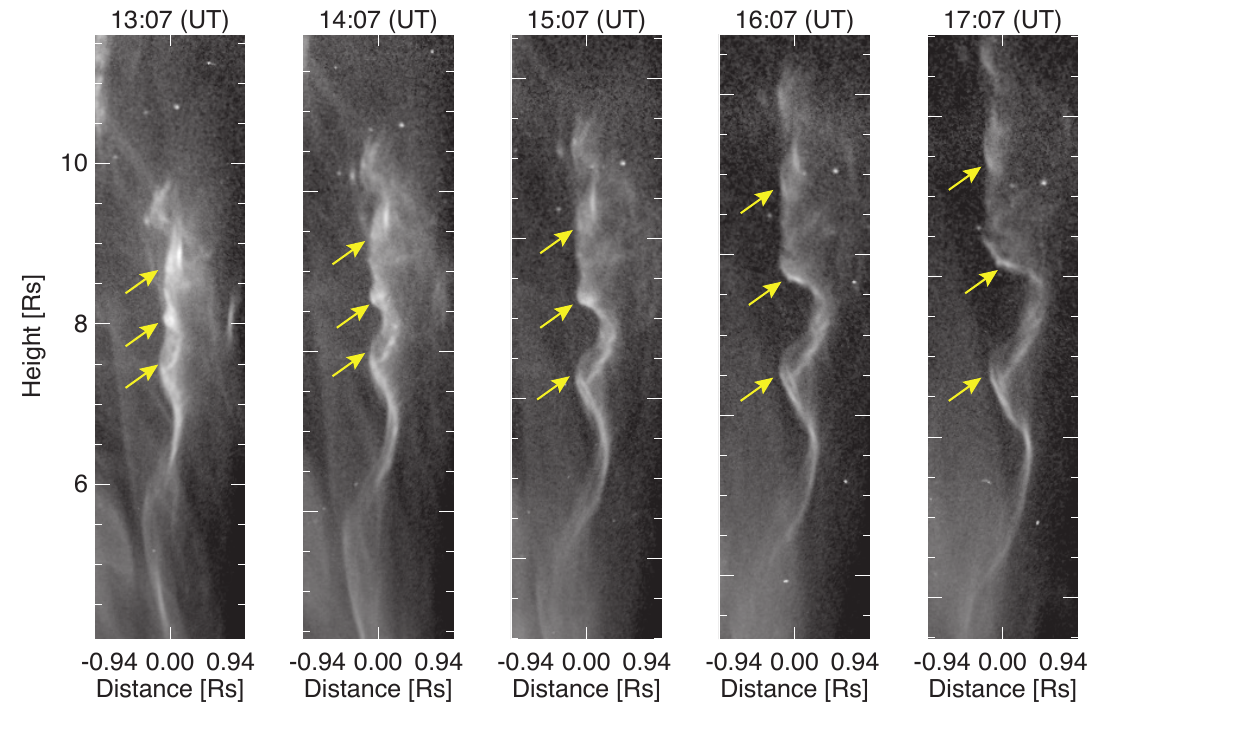}
\caption{Five snapshots from  the STEREO Cor 1 animation of the KHI feature shown Figure~\ref{slit_vr:fig}a at hourly intervals starting at 13:07 UT on 2024 February 16 showing the development of the KHI wave structure and the developed stage of KHI waves and vortices with their locations marked with yellow arrows.}
\label{khi_wave:fig}
\end{figure}

\begin{figure}[h]
\hspace{3.2cm}(a)\hspace{5.6cm}(b)\hspace{5.3cm}(c)
\begin{center}
\vspace{-0.3cm}\includegraphics[width=0.35\textwidth]{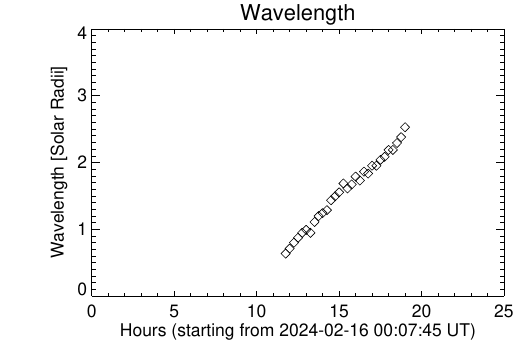}
\includegraphics[width=0.31\textwidth]{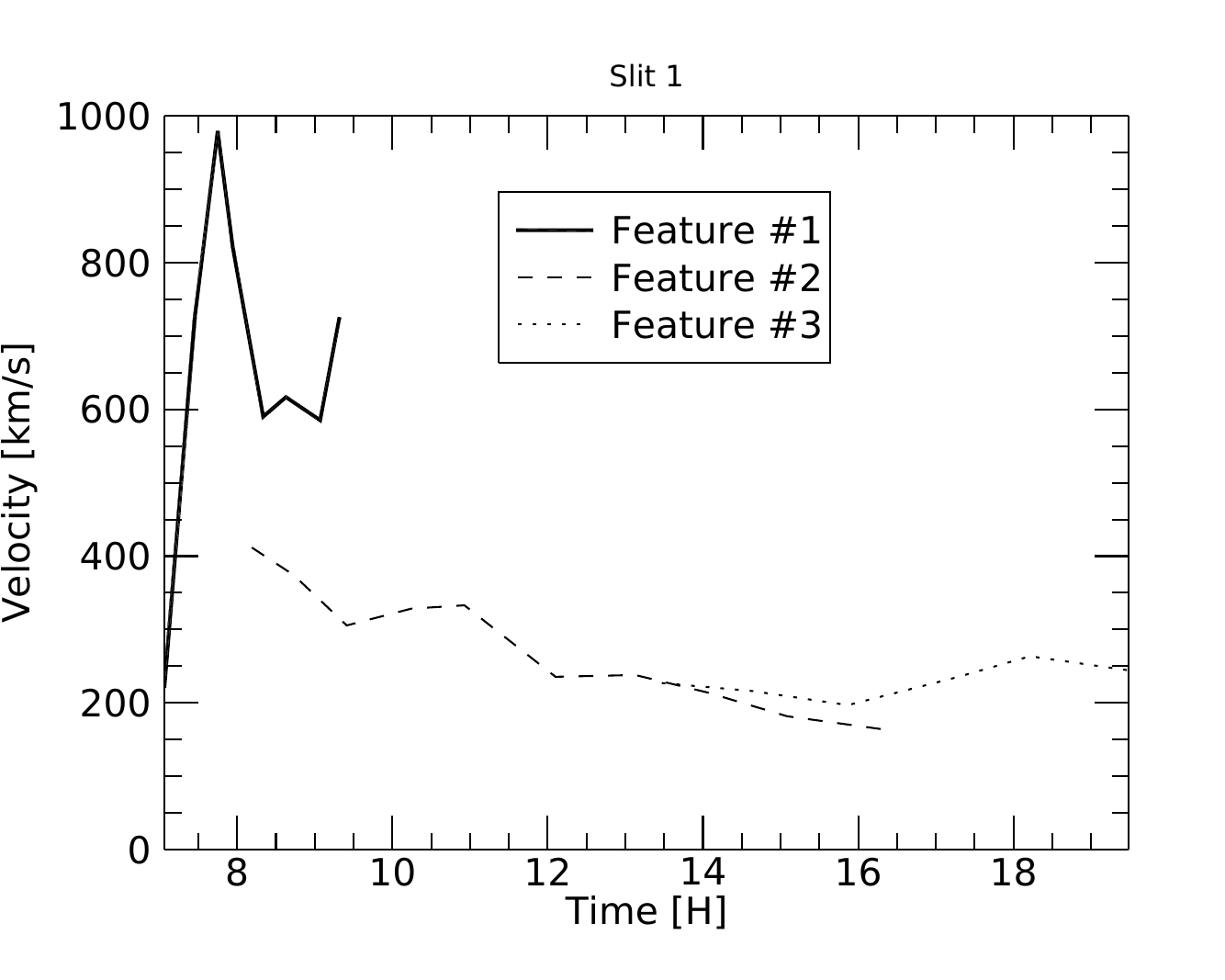}
\includegraphics[width=0.31\textwidth]{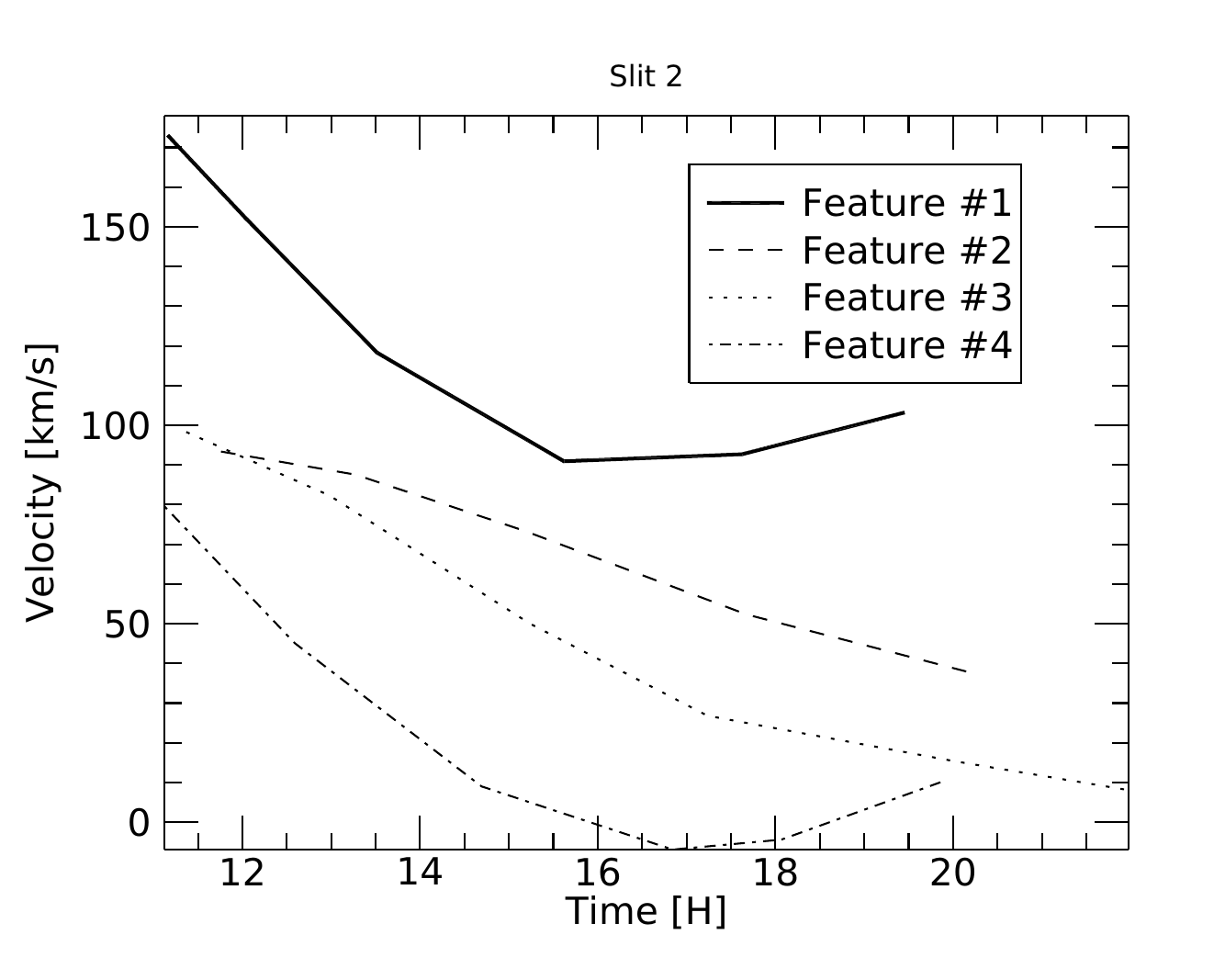}
\caption{(a) The temporal evolution of the wavelength of the KHI wave observed on 2024 February 16 by STEREO COR2 A, determined as shown at Slit 2 in Figure~\ref{slit_vr:fig}b. The temporal evolution of the velocities:  (b) The radial velocities of the features along Slit~1 (\#1 solid, \#2 dashes, \#3 dots) and (c) the evolution of velocities along Slit~2 (\#1 solid, \#2 dashes, \#3 dots, \#4 dot-dashes). The line styles are the same as in Figure~\ref{slit_vr:fig}.}
\label{wavelength-t:fig}
\end{center}
\end{figure}

In Figure~\ref{khi_wave:fig}  we show five snapshots from  the STEREO Cor 1 animation at hourly intervals starting at 13:07 UT on 2024 February 16 of the wave structure showing the developed stage of KHI waves and vortices with their locations marked with yellow arrows. The signatures of KHI waves and the associated vortices are discernible where the locations of the vortices marked with yellow arrows. The wavelength of the KHI wave increase from $\sim1R_s$ at 13.07UT to $\sim1.9R_s$ at 17:07UT (see, Figure~\ref{wavelength-t:fig}a).

In Figure~\ref{wavelength-t:fig}a the temporal evolution of the wavelength (in $R_s$ units) is shown (also, see Figure~\ref{slit_vr:fig}a and the corresponding animation). It is evident that the wavelength increases nearly linearly with time from $\sim0.6 R_s$ to $\sim2.5R_s$, where the peak wave amplitude occurs at $\sim16.5$ hours for a wavelength of $1.7R_s$. From the temporal evolution (see the online animation) it appears that there are two stages of evolution of the KHI wave: initial  growth of the KHI wave amplitude during the CME eruption,  followed by the damping the KHI wave amplitude at later times after the waning of the CME eruption and of the radial expansions of the wave structure.  

%4

The radial outflow velocities $\bf V_r$ of the features in the stationary Slit~1 are determined by evaluating the numerical derivatives of the space-time plot fits along the lines (indicated in colors) shown in Figure~\ref{wavelength-t:fig}b and c, respectively. The radial velocities $V_r$ for Slit 1 are shown in Figure~\ref{slit_vr:fig}a. It is evident that after $\sim7.5$ hours the $V_r$ reaches the peak value close to 1000 km s$^{-1}$, following by decrease to $\sim 600-700$ km s$^{-1}$. However, the second and third features have much lower velocities decreasing from $\sim400$ km s$^{-1}$ to the range of $150-200$ km s$^{-1}$, consistent with the local solar wind speed. The third feature corresponds to the flank of the second weaker CME.

The flow velocities of the features in Slit~2 are shown in Figure~\ref{slit_vr:fig}c. Since this slit is along the KHI wave feature, it is evident that the velocities are generally slower than along Slit~1. The first feature associated with the KHI wave is detected in the slit at about 10 hours after the onset of the main CME event, reaching 300 km s$^{-1}$, and slows down to $\sim160$ km s$^{-1}$ at $t\approx 16$ hours. The feature \#1 detected in the slit start with $V\approx170$ km s$^{-1}$ and decelerates to $\sim 100$ km s$^{-1}$, while the remaining features \#2-\#3 are all slow with velocities below $100$ km s$^{-1}$. The feature \#4 becomes nearly stationary at $t=16$ hours. Thus, due to the variation of the outflow speed between the features it is evident that there is significant initial acceleration and deceleration of the erupting material as well as significant shear velocity between the out-flowing material as demonstrated by the analysis of the two slits. Base on the observed CME feature velocities in Slit~1 in the range of $\sim$400-1000  km s$^{-1}$  and the KHI  wave feature velocities in Slit 2 in the range $\sim$0-170 km s$^{-1}$, the velocity shear can be estimated in the range $\sim$230-830 $^{-1}$ and is likely supe-Alfv\'{e}nic favorable for KHI growth (see the theoretical analysis below).

\vspace{0.5cm}\subsection{Alfv\'{e}n and Solar Wind Speeds from PSP Data}
\label{psp_av:sec}
\begin{figure}[h]
\begin{center}
(a)\hspace{8cm}(b)\\
\includegraphics[width=0.49\textwidth]{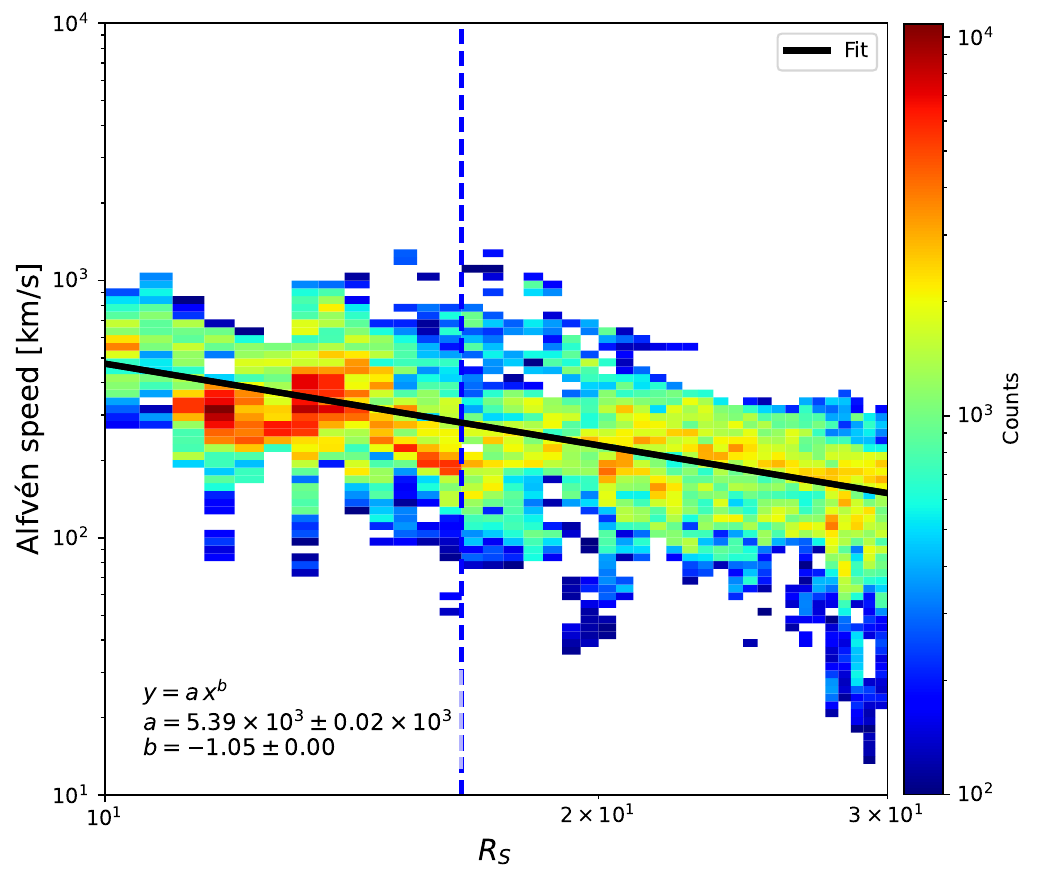}
\includegraphics[width=0.49\textwidth]{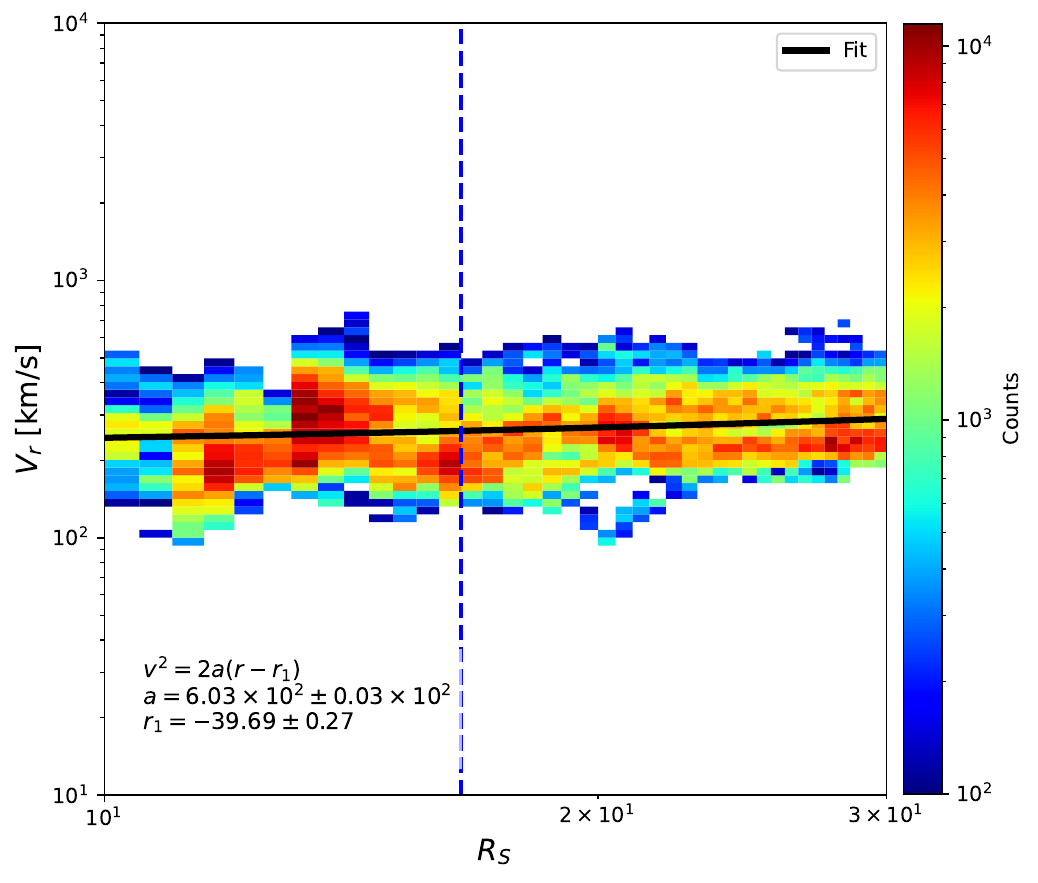}
\caption{The statistical distribution vs. heliocentric distance obtained from PSP in the range 10-30 $R_s$ obtained from  encounters 4-23. (a) The Alfv\'{e}n speed with power-law fit; (b) radial solar wind speed with \citet{She97} fit.  The dashed vertical line marks the mean Alfv\'{e}n
critical surface at 16.5 $R_s$ where $M_A = 1$. The color bar shows the number of counts. The black lines show the fits and the fit parameters are indicated on the panels.}
\label{psp_a_swv:fig}
\end{center}
\end{figure}

Before PSP era, the Alfv\'{e}n speed and the solar wind speed in the region of interest of CME eruption close to the Sun were determined indirectly and empirically using remote sensing \citep[e.g.,][]{She97,Col01,GY11}. Since the launch of PSP we now have direct in-situ measurements of the main solar wind parameters at perihelia as close as $9.86R_s$. Since there were no in-situ PSP observations during the exact time of the present event, we use the ensemble of PSP data for statistical determination of solar wind parameters in this region. In Figure~\ref{psp_a_swv:fig} the statistical distribution vs. heliocentric distance of the Alfv\'{e}n speed, $V_A$, and the radial solar wind speed, $V_r$, are shown. The values are deduced from PSP SPAN-I (that provided the proton velocities) and FIELDS instrument that provided the magnetic field magnitude in the distance range 10-30 $R_s$ obtained during  encounters 4-23. The local Alfv\'{e}n speed values are calculated by using the expression $V_A=B/(4\pi\rho)^{1/2}$ with the in-situ values of the magnetic field magnitude, $B$ from PSP/FIELDS, and electron density $\rho$ deduced from quasi-thermal noise (QTN) data. The power law fit to the Alfv\'{e}n speed $V_A$ is shown in Figure~\ref{psp_a_swv:fig}a with power $b=-1.05$.  The radial solar wind speed $V_r$ is fit with Equation~2 from \citet{She97}, i.e., $V_r^2=2a(r-r_1)$, the fit parameters are $a\approx6\times10^2$ and $r_1\approx-40$  shown in Figure~\ref{psp_a_swv:fig}b. The dashed vertical line marks the mean Alfv\'{e}n critical surface of the solar wind at 16.5 $R_s$ where $M_A = 1$. The color bar scale shows the number of counts for each quantity.

Since there were only few encounters at distances of $\sim10R_s$ we quote the statistical values of Alfv\'{e}n speed at 11.25$R_s$ using PSP data for encounters 4-23 (Yogesh et al. 2025 in prep.): the median value for the Alfv\'{e}n speed is 331 km s$^{-1}$, while p10 (10-percentile) value is 235 km s$^{-1}$, and the p90 value is 560 km s$^{-1}$. For $V_r$ at 11.25$R_s$ the p10 value was 126 km s$^{-1}$, the median value was 191 km s$^{-1}$, and p90 value  was 283 km s$^{-1}$. The general trends of PSP data in this distance range indicate that the above $V_A$ is near the upper limit, while $V_r$ is near the lower limit. These PSP-derived values of the Alfv\'{e}n speed are consistent with previous estimates and observations at $\sim10R_s$, such as \citet{Col01} that estimates $V_A=230$ km s$^{-1}$ at $10R_s$. 
%5

\section{Theoretical KHI Analysis}
\label{tkhi:sec}
The linear growth rate of KHI in the presence of magnetic field  for a discontinuous shear layer derived and used previously \citep[e.g.,][]{Cha61,Fra96,OT11} can be written as
\begin{eqnarray}
&&\gamma=\frac{1}{2}\left|\mbox{\bf k}\cdot\Delta\mbox{\bf V}\right|\left[1-(2V_A\hat{\bf k}\cdot\hat{\bf B})^2/(\hat{\bf k}\cdot\Delta \mbox{\bf V})^2\right]^{1/2},
\label{gam:eq}
\end{eqnarray}
where $\Delta V$ is the velocity jump across the interface, $\mbox{\bf B}$ is the magnetic field, $\mbox{\bf k}$ is the wavevector, the $\hat{}$ indicates a unit vector, and $V_A$ is the Alfv\'{e}n speed. The above expression provides an upper limit for a finite width velocity interface. It is evident that the parallel flow speed jump must exceed the local Alfv\'{e}n speed $V_A=B/(4\pi\rho)^{1/2}$, where $\rho$ is the density, so that the KHI could grow.

From observations we estimate $\Delta V\approx 435$ km s$^{-1}$, the wavevector magnitude is given by $k=2\pi/\lambda$, where the estimated wavelength  $\lambda\sim 0.6 R_s=420,000$ km. Assuming $\mbox{\bf k}\cdot \mbox{\bf B}\approx0$ we get the growth rate $\gamma\approx 0.0033$ s$^{-1}$ equal to $e$-folding growth time of $\sim~5$ minutes and the KHI nonlinear saturation time is several $e$-folding growth times. The estimated growth rate is in qualitative agreement with the observed growth and evolution of the KHI wave due to the  super-Alfv\'{e}nic CME eruption on 2024 February 16. For the specific conditions $\mbox{\bf k}\parallel \Delta\mbox{\bf V}\parallel \mbox{\bf B}$ the  growth rate can be estimated from the simplified expression 
\begin{eqnarray}
&&\gamma=\frac{1}{2}\left|k\Delta V\right|\left[1-(2V_A)^2/(\Delta V)^2\right]^{1/2}.
\label{gam_par:eq}
\end{eqnarray}
%{\color{blue}EH: can we show in a graph a comparison between the theoretical estimation of (2) to the observed growth?}

It is also useful to calculate the wave-number normalized growth-rate and the portion of the $\pi$-angle (($\Omega_{\pi}$, hereafter)  radially away from the Sun that is KHI-unstable and assuming that  $\hat{\bf{k}}$, $\hat{\bf{B}}$, and $\Delta {\bf{V}}$ are in the same plane and $\Delta {\bf{V}}$ is along $x$-axis. The angle between $\hat{\bf{k}}$ above (below) the $x$-axis and $\Delta {\bf{V}}$ is $\theta$ (-$\theta$), and the angle between $\hat{\bf{k}}$ and ${\bf{B}}= \alpha$  between [0,$\pi$/2] rad, and 2$\theta$+$\alpha$ between [0,$-\pi/$2] rad. To evaluate this we express Equation~\ref{gam:eq} in the following form: 

\begin{eqnarray}
&&\left(\frac{\gamma}{k}\right)^2=\frac{1}{4}\left|\Delta \mbox{V}\cos(\theta)\right|^2\left[1-(2V_A\cos(\alpha))^2/(\Delta \mbox{V}\cos(\theta))^2\right].
\label{gam_k:eq}
\end{eqnarray}

Since the values  of the magnetic field, the Alfv\'{e}n speed, ${\bf k}$ and angle with ${\bf B}$ are not directly available from the present observations, we next solve this equation with a Monte-Carlo method by creating large number $N$ values of $\theta$ and $\alpha$- angles spanning the range  [0, $\pi$/2] rad (above $x$-axis) and then repeat for the range [0,-$\pi/$2] rad where $\hat{\bf k}$ is below $x$-axis (but $\hat{\bf{B}}$ is above $x$- axis) and collecting all vectors for which $(\gamma/k)^2 > 0$. The number $N$ needs to be large such that results converge (similar method has been used in, e.g., \citet{burk20,Nyk21b,Nyk24} where the whole solid angle was searched due to additional  degrees of freedom in single point in-situ data). The $V_A$ statistical values from PSP data (p10, median, and p90) and $\Delta V=435$ km s$^{-1}$ yield the unstable fractions of the solid angle $\Omega_{\pi} =$ 45$\%$, 31$\%$, and 18$\%$, respectively for $N=10^6$ and the results converge when running several times and with increased $N$. The maximum growth rate naturally corresponds to the situation when $\hat {\bf k}$ and $\hat {\bf{B}}$ are orthogonal, and the velocity shear is parallel to  $\hat {\bf k}$. This yields $\gamma/k=$ 217 km/s, the same result as determined above. We note that the above idealized theoretical analysis in simplified geometry does not take into account the possible helical structure of the CME and the potential effects of this structure on the growth rate of KHI.

\section{Discussion and Conclusion}
\label{disc:sec}
Using STEREO COR2 and SOHO/LASCO C2/C3 coronagraph imaging observations we find evidence for KHI in large amplitude wave growth and dissipation in a narrow elongated coronal feature likely associated with partially erupted prominence material at distances of $\sim 6-14 R_s$ associated with the CME on 2024 February 16, commenced at 7:12 UT. The observed KHI wave growth event is rarely detected, with  KHI wave features detected in a handful events at this range of radial distances in streamers associated with CME events \citep[e.g.,][]{Fen13,Pao24}. This is the first analysis of a large-scale large amplitude KHI wave event associated with super-Alfv\'{e}nic CME through all stages of growth and dissipation. Our analysis provides the detailed wave properties, as well as velocities of the erupting CME and streamer features close to the Sun. It is apparent from coronagraph observations that the main CME is followed by a secondary weaker CME several hours later, that likely facilitated further growth of KHI waves in the adjacent streamer feature due to the combined prolonged eruption stage.

Since the growth of KHI can be suppressed by parallel magnetic field, the velocity shear must exceed the local (at $\sim 6-14 R_s$) parallel Alfv\'{e}n speed to allow growth of KHI \citep{Cha61}. In order to evaluate the ambient stability conditions associated with the CME, we estimate the local Alfv\'{e}n speed and the solar wind speed at the observed KHI wave distances using PSP statistical perihelia data from encounters 4-23. We find that PSP perihelia data of $V_A$ and $V_r$ are in qualitative agreement with previous empirical and remote sensing estimates of these quantities. Since the exact parameters of the solar corona, such as magnetic field strength and orientation are not well known, we explore theoretically  the parameter space  that allows KHI growth using the Monte Carlo method. As expected, the fastest growth of KHI takes place for orthogonal wavenumber with respect to the magnetic field we find that significant fraction of possible orientations can results in KHI onset. Our analysis shows that the KHI can grow at a distance beyond 6$R_s$ where the CME propagation speed is super-Alfv\'{e}nic, leading to sufficiently large velocity shear with the ambient streamer features, enabling the growth of KHI. The subsequent damping of KHI wave is consistent with the waning stage of the CME eruption, and associated decrease in local velocity shear down to sub-Alfv\'{e}nic values.

The present CME event is favorable for observing the instability growth in its trailing edge, likely due to the radial elongation of magnetic-field lines, reduced plasma density following the passage of the CME, and enhanced velocity and magnetic-field shear along the developing interface with the adjacent quasi-stationary bright elongated streamer-like structure. As was discussed in several previous studies and demonstrated here, the KHI can play an important role in the energy transfer between the large scale CME eruptions and the smaller scale coronal structures, that can eventually result in coronal and solar wind heating as well in the diagnostic of magnetic field and flows in the lower corona. The present results improve  the understanding of the nonlinear large-scale evolution of KHI  by using the combination of remote sensing with new in-situ data, together with theoretical analysis to study the observed large-scale KHI wave in previously unexplored region close to the Sun, where KHI can play an important role in the energetics of the solar corona.

%% Please use the acknowledgment and contribution environments. This will 
%% be anonomyized when the "anonymous" style option is used. 
\begin{acknowledgments}
The authors L.O. and Y. acknowledge support by NASA grant 80NSSC24K0724, NSF grant AGS-2300961 as well as from NASA GSFC through Cooperative Agreement 80NSSC21M0180 to Catholic University of America, Partnership for Heliophysics and Space Environment Research (PHaSER). L.O. and K.N. acknowledge support by NASA Internal Scientist Funding Model (ISFM) at GSFC. O.K. acknowledges support by Center of Absorption in Science, Ministry of Immigration and Absorption of Israel for her work at Tel Aviv University. 
Y. acknowledge support by NASA Heliophysics Guest Investigator Grant 80NSSC23K0447 and the College of Liberal Arts and Sciences at the University of Iowa. K.N. also acknowledges the support from NASA Living with a Star grant 80NSSC23K0899.
\end{acknowledgments}

%\begin{contribution}
%%This section gives authors the space to recognize author contributions. The text inside this environment is NOT counted towards the total word quanta. At a minimum, manuscripts are expected to include this text:

%All authors contributed equally to the Terra Mater collaboration.

%% But authors are expected to provide more specific details, e.g. 
%%
%%SC was responsible for writing and submitting the manuscript.
%%WWM came up with the initial research concept and edited the manuscript.
%%OTS obtained the funding and edited the manuscript.
%%EBF provided the formal analysis and validation. He also edited the manuscript.
%%GEH Supervised the undergraduates, wrote the software and administers the project github and Zenodo repositories.
%%
%% Authors can use the Contributor Role Taxonomy (CRediT) at
%% https://credit.niso.org
%% for ideas on how write a good statement tailored to their needs.

%\end{contribution}

%% To help institutions obtain information on the effectiveness of their 
%% telescopes the AAS Journals has created a group of keywords for telescope 
%% facilities.
%
%% Following the acknowledgments section, use the following syntax and the
%% \facility{} or \facilities{} macros to list the keywords of facilities used 
%% in the research for the paper.  Each keyword is check against the master 
%% list during copy editing.  Individual instruments can be provided in 
%% parentheses, after the keyword, but they are not verified.
\facilities{STEREO/COR A, SOHO/LASCO, SDO/AIA, PSP/SPAN-I, PSP/FIELDS}

%% Similar to \facility{}, there is the optional \software command to allow 
%% authors a place to specify which programs were used during the creation of 
%% the manuscript. Authors should list each code and include either a
%% citation or url to the code inside ()s when available.

%% Appendix material should be preceded with a single \appendix command.
%% There should be a \section command for each appendix. Mark appendix
%% subsections with the same markup you use in the main body of the paper.
%%
%% Each Appendix (indicated with \section) will be lettered A, B, C, etc.
%% The equation counter will reset when it encounters the \appendix
%% command and will number appendix equations (A1), (A2), etc. The
%% Figure and Table counter will not reset.

%% For this sample we use BibTeX plus aasjournalv7.bst to generate the
%% the bibliography. The sample7.bib file was populated from ADS. To
%% get the citations to show in the compiled file do the following:
%%
%% pdflatex sample7.tex
%% bibtext sample7
%% pdflatex sample7.tex
%% pdflatex sample7.tex

\bibliography{ofman_khi}
\bibliographystyle{aasjournalv7}

%% This command is needed to show the entire author+affiliation list when
%% the collaboration and author truncation commands are used.  It has to
%% go at the end of the manuscript.
%\allauthors

%% Include this line if you are using the \added, \replaced, \deleted
%% commands to see a summary list of all changes at the end of the article.
%\listofchanges

\end{document}